\documentclass[twocolumn,showpacs,preprintnumbers,amsmath,amssymb]{revtex4}

\usepackage{graphicx}
\usepackage{dcolumn}
\usepackage{bm}

\begin{document}


\title{Temperature evolution of spin accumulation detected electrically in a nondegenerated silicon channel}

\author{Y. Ando,$^{1,2}$ K. Kasahara,$^{1}$ S. Yamada,$^{1}$ Y. Maeda,$^{1}$ K. Masaki,$^{1}$ Y. Hoshi,$^{3}$ K. Sawano,$^{3}$ M. Miyao,$^{1}$ and K. Hamaya$^{1,4}$\footnote{E-mail: hamaya@ed.kyushu-u.ac.jp}}

\affiliation{$^{1}$Department of Electronics, Kyushu University, 744 Motooka, Fukuoka 819-0395, Japan}%
\affiliation{$^{2}$INAMORI Frontier Research Center, Kyushu University, 744 Motooka, Fukuoka 819-0395, Japan}%
\affiliation{$^{3}$Advanced Research Laboratories, Tokyo City University, 8-15-1 Todoroki, Tokyo 158-0082, Japan}
\affiliation{$^{4}$PRESTO, Japan Science and Technology Agency, Sanbancho, Tokyo 102-0075, Japan}%

%

\date{\today}
\begin{abstract}
We study temperature evolution of spin accumulation signals obtained by the three-terminal Hanle effect measurements in a nondegenerated silicon channel with a Schottky-tunnel-barrier contact. We find the clear difference in the temperature-dependent spin signals between spin-extraction and spin-injection conditions. In a spin-injection condition with a low bias current, the magnitude of spin signals can be enhanced despite the rise of temperature. For the interpretation of the temperature-dependent spin signals, it is important to consider the sensitivity of the spin detection at the Schottky-tunnel-barrier contact in addition to the spin diffusion in Si. 
\end{abstract}
\maketitle

\section{INTRODUCTION}
Electrical detection of the spin accumulation is one of the important technologies to achieve semiconductor-based spintronic applications.\cite{Dery1} Though excellent studies of these technologies were reported for GaAs-based devices, almost all the spin-related phenomena were observed below room temperature.\cite{Lou,Ciorga,Salis,Uemura} In general, it has been understood that the decay of the spin signals originates from the spin-flip scattering induced by the spin-orbit interaction, hyperfine interaction between electrons and nuclei, and so on.\cite{Kikkawa} To minimize such intrinsic factors, silicon-based spintronic technologies have been proposed and developed,\cite{Min,Igor1,Appelbaum,Jonker1,Ando,Sasaki1} and room-temperature detections of the spin-dependent signals have been reported recently.\cite{Jansen,Suzuki,Ando3} 

The temperature-dependent spin-related phenomena in Si-based devices were explored experimentally.\cite{Jansen,Appelbaum2,Sasaki2,Suzuki} For nondoped Si channels,\cite{Appelbaum2} the relaxation of the injected spins can be explained by Yafet's $T^{-5/2}$ power law, indicating that the decay of the spin signals is attributed to the spin-flip scattering in the channel. For heavily doped Si (degenerated Si) channels,  on the other hand, the decrease in the spin polarization of the injected spins is one of the main factors for the decay of spin signals.\cite{Sasaki2} However, there is no study of the temperature-dependent  spin signals for nondegenerated Si channels. 

In this article, we report on temperature evolution of spin signals by measuring the three-terminal Hanle effect in the lateral device with a nondegenerated silicon channel. In our device with a CoFe/Si Schottky-tunnel-barrier contact, a clear difference in the temperature-dependent spin signals between spin-extraction and spin-injection conditions is observed. Under a certain condition, the sensitivity of the spin detection can contribute dominantly to the magnitude of spin signals detected. 

\section{Samples and measurements}
The CoFe epitaxial layer was grown on Si(111) by low-temperature molecular beam epitaxy (LT-MBE) at 60 $^{\circ}$C and the CoFe/Si interface was atomically flat.\cite{Maeda} A three-terminal lateral device (channel thickness $\sim$ 100 nm, carrier density $ \sim$ 6.0 $\times$ 10$^{17}$ cm$^{-3}$) with one single-crystalline Co$_{60}$Fe$_{40}$ contact and two AuSb ohmic ones with lateral dimensions of 10 $\times$ 200 $\mu$m$^{2}$ and 100 $\times$ 200 $\mu$m$^{2}$, respectively, was fabricated by using conventional processes with photolithography, Ar$^{+}$ ion milling, and reactive ion etching.\cite{Ando,Ando2,Ando3} The schematic diagram of the device structure is shown in the inset of Fig. 1(e). The distance between CoFe and AuSb is about $\sim$ 50 $\mu$m. To achieve tunneling conduction through the high-quality CoFe/Si interface, we inserted Sb $\delta$-doped $n^{+}$-Si layer (Sb $\sim$ 5$\times$10$^{19}$ cm$^{-3}$) with a thickness of 5 nm between the epitaxial CoFe layer and $n$-Si channel.\cite{Ando,Ando2,Ando3} As a result, we obtained tunneling conduction having nonlinear $I - V$ characteristics through the interface and the rectification is quite small, as shown in Fig. 1(a). Therefore, we can regard the fabricated CoFe/$n^{+}$-Si/$n$-Si contact for spin injection and extraction as a Schottky-tunnel-barrier contact. Hanle-effect measurements with a three-terminal geometry were performed by a dc method at 40 $\sim$ 300 K. In the measurements, a small magnetic field perpendicular to the plane, $B_\text{Z}$, was applied after the magnetic moment of the CoFe contact aligned parallel to the plane along the long axis of the contact.  

\section{Results and Discussion}  
Spin accumulation in semiconductor channels can be detected electrically by measuring three-terminal voltage changes ($\Delta$$V_\text{3T}$) via Hanle-type spin precessions.\cite{Lou,Jansen,Ando2,Ando3} Figures 1(c) and 1(d) show representative $\Delta$$V_\text{3T}$-$B_\text{Z}$ curves for $I  =$ -1.0 and 1.0 $\mu$A at 40 K. The electrons are injected into and extracted from, respectively, the Si conduction band for reverse ($I <$ 0) and forward ($I >$ 0) biases, where quadratic background voltages depending on $B_\text{Z}$ are subtracted from the raw data. We note that clear Hanle-effect signals were observed for both bias-current conditions but the sign of the voltage change is opposite. Such sign reversal was seen only when the polarity of $I$ was switched. These curves were fitted with a simple Lorentzian function,\cite{Jansen} $\Delta$$V_\text{3T}$($B_\text{Z}$) $=$ $\Delta$$V_\text{3T}(0)$/[1+($\omega_\text{L}$$\tau_\text{S}$)$^{2}$], where $\omega_\text{L} =$ $g\mu_\text{B}$$B_\text{Z}$/$\hbar$ is the Larmor frequency, $g$ is the electron $g$-factor ($g =$ 2), $\mu_\text{B}$ is the Bohr magneton, $\tau_\text{S}$ is the lower limit of spin relaxation time. The fitting results are denoted by the red solid curves. The $\tau_\text{S}$ values for $I  =$ -1.0 and 1.0 $\mu$A are estimated to be $\sim$ 4.37 and $\sim$ 2.05 nsec, respectively. The precise origin of the difference in $\tau_\text{S}$ between $I  =$ -1.0 and 1.0 $\mu$A is unclear yet, but we can consider a possible difference in the relative position of the spin accumulation in $n$-Si to the $n^{+}$-Si layer, as illustrated in Fig. 1(b). There may be some differences in the influence of the $n^{+}$-Si layer on spin relaxation between spin injection ($I <$ 0) and extraction ($I >$ 0) conditions.
\begin{figure}[t]
\includegraphics[width=7.5cm]{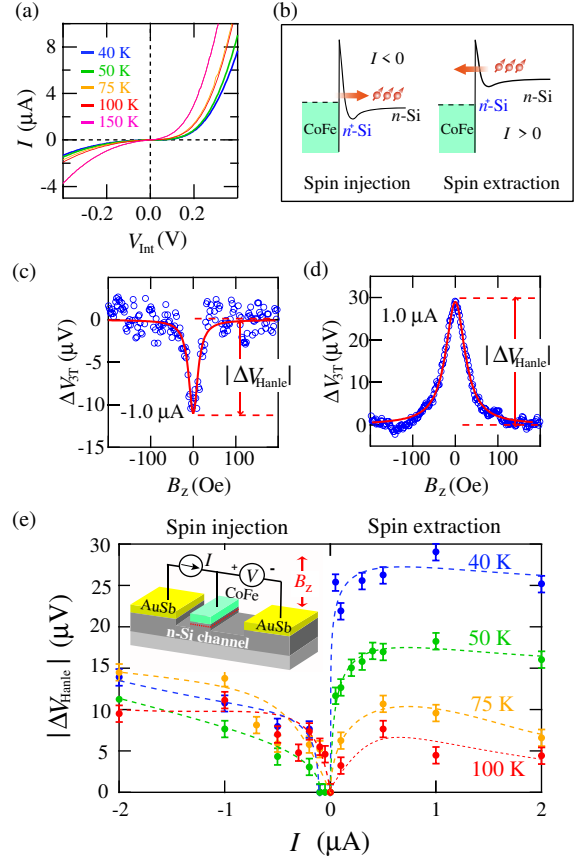}
\caption{(Color online) (a) $I - V_\text{Int.}$ characteristics for various temperatures. (b) Schematic illustration of spin accumulation in spin-injection and spin-extraction conditions. $\Delta$$V_\text{3T}$-$B_\text{Z}$ curves at (c) $I$$ =$ -1.0 and (d) 1.0 $\mu$A at 40 K, and fitting results with Lorentzian function (solid curves).\cite{Jansen} (e) The magnitude of the observed spin accumulation signals, $|$$\Delta$$V_\text{Hanle}$$|$, as a function of bias current $I$ for various temperatures. }
\end{figure}

We hereafter focus on the magnitude of $\Delta$$V_\text{3T}$, $|\Delta$$V_\text{Hanle}|$, for various temperatures. Figure 1(e) displays $|\Delta$$V_\text{Hanle}|$ versus bias current $I$ (-2.0 $\mu$A  $\le I \le$ 2.0 $\mu$A) at 40, 50, 75, and 100 K. We can see clear asymmetric variation in $|\Delta$$V_\text{Hanle}|$ with respect to the polarity of $I$.\cite{ref} For $I >$ 0, $|\Delta$$V_\text{Hanle}|$ decreases with increasing temperature in all $I$ region while, for $I <$ 0, complicated variations in $|\Delta$$V_\text{Hanle}|$ are seen, particularly, in -1.0 $\mu$A $\le$ $I \le$ 0 $\mu$A. We can see that $|\Delta$$V_\text{Hanle}|$ at 100 K is larger than that at 50 K. Concentrating on this interesting phenomenon, we further explored temperature-dependent $|\Delta$$V_\text{Hanle}|$ for various $I$ in detail. Figures 2(a) and 2(b) show $|\Delta$$V_\text{Hanle}|$ as a function of temperature for spin-extraction ($I >$ 0) and spin-injection ($I <$ 0) conditions, respectively. The features observed are summarized as follows: (i) Temperature evolution of $|\Delta$$V_\text{Hanle}|$ shows a clear difference between spin-extraction ($I >$ 0) and spin-injection ($I <$ 0) conditions. (ii) $|\Delta$$V_\text{Hanle}|$ can be detected even at room-temperature with a large injection current of - 20 $\mu$A in $I <$ 0 [see the inset of Fig. 2(b)], whereas $|\Delta$$V_\text{Hanle}|$ is markedly reduced with increasing temperature and disappears at 250 K irrespective of $I$ in $I >$ 0. $\tau_\text{S}$ for $I =$ -20 $\mu$A at room temperature can be estimated to be 1.36 nsec. The $\tau_\text{S}$ value decreases with increasing temperature from 3.02 to 1.36 nsec, consistent with the feature observed in the heavily doped Si.\cite{Sasaki2,Suzuki} (iii) In $I =$ -0.1 and -1.0 $\mu$A we can see the partial increases in $|\Delta$$V_\text{Hanle}|$ despite the rise of temperature [see arrows in Fig. 2(b)]. 
\begin{figure}[t]
\includegraphics[width=8.5cm]{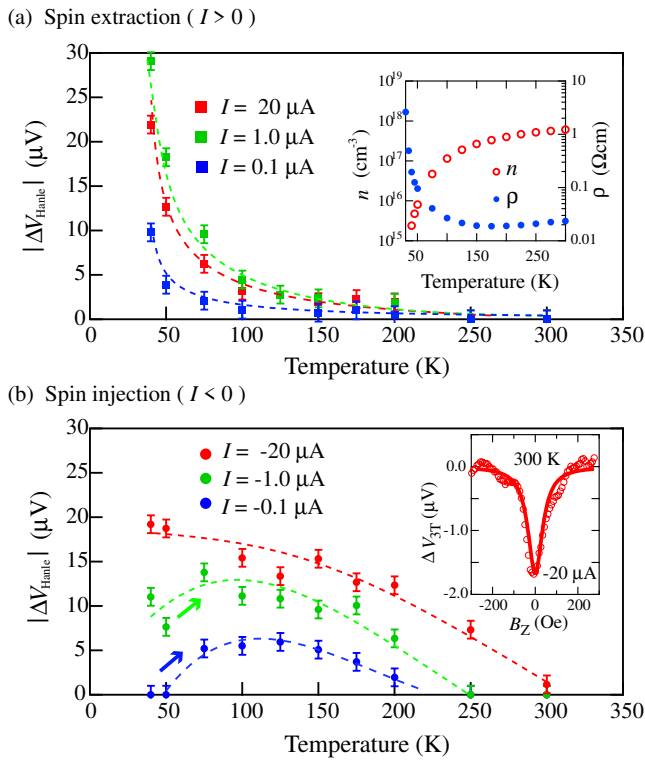}
\caption{(Color online) Temperature dependence of $|$$\Delta$$V_\text{Hanle}$$|$ for (a) spin-extraction ($I >$ 0) and (b) spin-injection ($I <$ 0) conditions. The inset of (a) shows carrier density and resistivity as a function of temperature for the Si channel used in this study. The inset of (b) displays room-temperature Hanle-effect signal at $I =$ -20 $\mu$A and the fitting curve with Lorentzian function.}
\end{figure} 

To examine these phenomena, first of all, temperature dependences of resistivity ($\rho_\text{Si}$) and carrier density ($n$) were shown in the inset of Fig. 2(a), where $\rho_\text{Si}$ and $n$ were measured directly by four-terminal transport and Hall effect measurements. Both $\rho_\text{Si}$ and $n$ have strong temperature dependence in $T <$ 100 K, nearly consistent with the feature of a general nondegenerated Si.\cite{Pearson} In this context, we should consider the component of $|\Delta$$V_\text{Hanle}|$ associated with changing $\rho_\text{Si}$ on the basis of the spin diffusion model.\cite{Fert} Since the spin-related voltage changes in Si are proportional to $\rho_\text{Si}$$\times$$\lambda_\text{Si}$,\cite{Fert} the variation in $\rho_\text{Si}$ with changing temperature can affect $|\Delta$$V_\text{Hanle}|$ in our measurements. The detail is discussed later.

Next, in Fig. 3 we examined $|\Delta$$V_\text{Hanle}|$ versus $V_\text{Int.}$, i.e., the bias voltage at the CoFe/Si interface for various temperatures. The $|\Delta$$V_\text{Hanle}|$ in $V_\text{Int.} >$ 0 has an evident maximum value at a certain $V_\text{Int.}$, where $V_\text{Int.} >$ 0 indicates spin-extraction conditions ($I >$ 0). This means that there is the most sensitive $V_\text{Int.}$ value for detecting spin-accumulation signals in $I >$ 0. We would like to define the detectability for the spin accumulation signals as the sensitivity of the spin detection at the biased contact. In general, the sensitivity of the spin detection at the Schottky-tunnel-barrier contacts has already been discussed for ferromagnet/semiconductor lateral devices.\cite{Lou,Salis,Jansen2,Smith,Crooker}  The observed feature also implies that we should consider the sensitivity of the spin detection for our CoFe/Si Schottky-tunnel-barrier contacts. However, when we focus on the temperature evolution of $|\Delta$$V_\text{Hanle}|$ at a certain $V_\text{Int.}$, the feature of the reduction in $|\Delta$$V_\text{Hanle}|$ with increasing temperature is relatively simple in $V_\text{Int.} >$ 0. Namely, we do not need to consider the change in the spin-detection sensitivity with temperature evolution in $I >$ 0. On the other hand, there is almost no clear correlation between $|\Delta$$V_\text{Hanle}|$ and $V_\text{Int.}$ in $V_\text{Int.} <$ 0, where $V_\text{Int.} <$ 0 indicates spin-injection conditions ($I <$ 0). We find that in -0.1 V $\le$ $V_\text{Int.}$ $\le$ 0 V $|\Delta$$V_\text{Hanle}|$ at 75 and 100 K is higher than that at 40 and 50 K. Namely, for $I <$ 0, we should also consider the sensitivity of the spin detection with temperature evolution.
\begin{figure}[t]
\includegraphics[width=7.5cm]{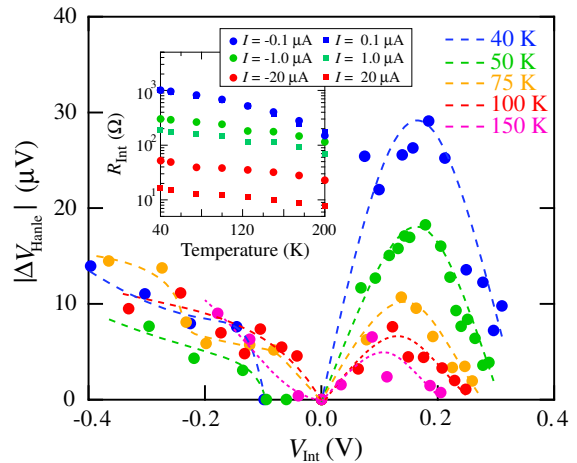}
\caption{(Color online) $|$$\Delta$$V_\text{Hanle}$$|$ as a function of $V_\text{Int.}$ for various temperatures. The insets show temperature-dependent $R_\text{Int.}$ for various $I$ conditions.}
\end{figure}  

Here we also concentrate on the variation in the interface resistance ($R_\text{Int.}$). The inset of Fig. 3 shows $R_\text{Int.}$ as a function of temperature for various $I$ conditions. Since the $I - V$ characteristics of our device are nonlinear features, the $R_\text{Int.}$ values can be varied with changing $I$. It should be noted that $R_\text{Int.}$ slightly decreases with increasing temperature for all $I$ conditions. Hence $V_\text{Int.}$ can also slightly be changed by the variation in temperature even if we use the same $I$ condition for the spin injection or extraction.  

Considering these features described so far, we discuss the temperature evolution of $|\Delta$$V_\text{Hanle}|$. Under spin-extraction conditions ($I >$ 0) [see Fig. 4(a)], we do not have to consider the temperature-dependent sensitivity of the spin detection at a certain $V_\text{Int.}$, as described in previous paragraph. We explain the detailed picture as follows. When the electrons tunnel from the Si conduction band into the spin-polarized empty states in CoFe at low temperature (e.g. 40 K) under $V_\text{Int.}$ ($\mu_\text{Int.}$), the spin accumulation ($\Delta$$\mu$) can be formed in the Si conduction band, as shown in the left figure of Fig. 4(a). In this situation, the spin-dependent tunneling of electrons is dominant at the Fermi level (see-dashed pink line), leading to the detection of $\Delta$$\mu$. 
As the temperature rises [e.g. 100 K, see right figure of Fig. 4(a)], $V_\text{Int.}$ is reduced, resulting in a small decrease in $\mu_\text{Int.}$. Simultaneously, $n$ is steeply enhanced from $n \sim$10$^{15}$ to $\sim$10$^{17}$ cm$^{-3}$, causing the rapid decrease in $\rho_\text{Si}$. As a result, though the $\Delta$$\mu$ value decreases markedly based on the spin diffusion model,\cite{Fert} there is almost no difference in the basic situation for the detection of $\Delta$$\mu$ after the rise of temperature. In other words, we can detect $\Delta$$\mu$ induced by the spin extraction irrespective of temperature because the spin-dependent tunneling of electrons is maintained. Therefore, there is the sensitivity of the spin detection. In this case, since $\Delta$$\mu$ in the Si conduction band is related directly to the observed $|\Delta$$V_\text{Hanle}|$ values, the change in $\rho_\text{Si}$ can dominantly affect the $|\Delta$$V_\text{Hanle}|$ values on the basis of the spin diffusion model. In Fig. 2(a) we can find the steep and gradual decreases in $|\Delta$$V_\text{Hanle}|$ in $T <$ 100 K and $T >$ 100 K, respectively. Since $\rho_\text{Si}$ in $T >$ 100 K is almost constant, the contribution of $\rho_\text{Si}$ to the decrease in $|\Delta$$V_\text{Hanle}|$ can be ignored in $T >$ 100 K. We infer that, for spin-extraction conditions, the features in $T <$ 100 K and $T >$ 100 K are dominated by the reduction in $\rho_\text{Si}$ and $\lambda_{\rm Si}$ (increase in the spin-flip scattering in Si), respectively. It seems that the change in $|\Delta$$V_\text{Hanle}|$ influenced by the spin-flip scattering is relatively small in the device with a nondegenerated Si channel. 
\begin{figure}[t]
\includegraphics[width=8.5cm]{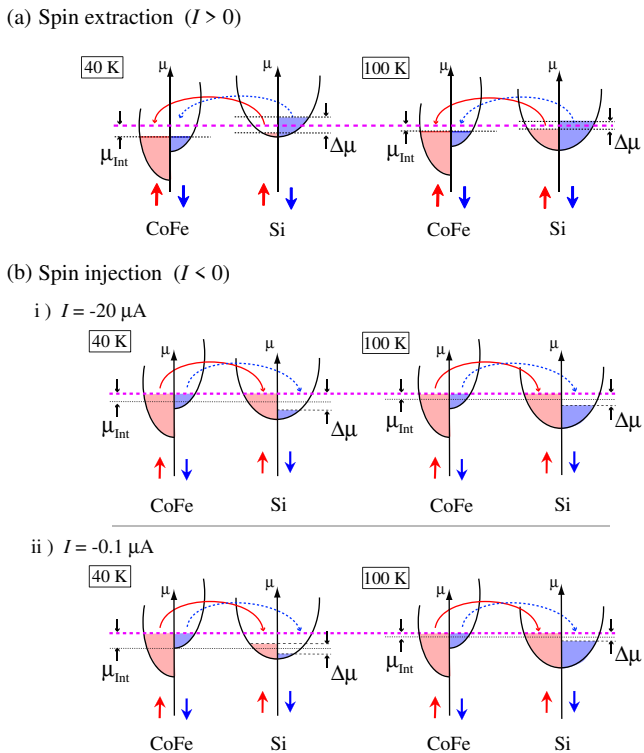}
\caption{(Color online) Schematic diagrams of the change in the spin accumulation signals for various temperatures. (a) Spin-extraction conditions at 40 and 100 K and (b) spin-injection conditions at 40 and 100 K with a i) high and ii) low bias current.}
\end{figure} 

Under spin-injection conditions ($I <$ 0), by contrast, we have to consider two different cases whether the injection current is relatively large or not, as shown in Fig. 4(b). When the large injection current (e.g. $I =$ -20 $\mu$A) is used, the relatively large $\Delta$$\mu$ can be demonstrated.\cite{Ando2} Thus $\Delta$$\mu$ induced in the Si conduction band can reach the Fermi level of the spin-polarized tunneling electrons at any temperatures. When the temperature rises, $\Delta$$\mu$ is merely reduced (middle figures) by the same mechanism of Fig. 4(a). Namely, we can still obtain the sensitivity of the spin detection. On the other hand, if the injection current is relatively low (e.g. $I =$ -0.1 $\mu$A) at 40 K, $\Delta$$\mu$ induced in the Si conduction band is quite small. Since the relatively large $\mu_\text{Int.}$ is applied in the low bias current regime at 40 K, we can understand that the position of the quasi Fermi level of the spin-polarized electrons in the Si conduction band is quite lower than the Fermi level (dashed pink line) of the spin-polarized tunneling electrons from CoFe. In this situation, the spin-polarized electrons can tunnel into the unpolarized states (unoccupied states) in the Si conduction band. Thus we could not obtain $\Delta$$\mu$ as a consequence of the $|\Delta$$V_\text{Hanle}|$ measurement at $I =$ -0.1 $\mu$A and 40 K. This means that there is no sensitivity of the spin detection. The similar situations have already been reported for Fe/GaAs devices previously,\cite{Lou} and we have also reported the above features in detail for the CoFe/Si devices.\cite{Ando2} 

Why we can see that $|\Delta$$V_\text{Hanle}|$ appears at 100 K for $I =$ -0.1 $\mu$A despite the rise of temperature? Shown in the right figure of the bottom in Fig. 4(b) is a possible schematic illustration of the interpretation of the spin detection. As previously described, when the temperature rises up to 100 K, $n$ is enhanced from $n \sim$ 10$^{15}$ to 10$^{17}$ cm$^{-3}$, causing the rapid decrease in $\rho_\text{Si}$. As a result, $\Delta$$\mu$ is also reduced on the basis of the spin diffusion model.\cite{Fert} Simultaneously, since $V_\text{Int.}$ also decreases with decreasing $R_\text{Int.}$, the reduction in $\mu_\text{Int.}$ also occurs. Accordingly, the Fermi level of the spin-polarized tunneling electrons is located on $\Delta$$\mu$ induced in the Si conduction band,
leading to the appearance of the sensitivity of the spin detection. This can be considered to be the same situation for $I =$ -20 $\mu$A at 40 K shown in the left figure of Fig. 4(b). This is a possible mechanism of the appearance of the spin accumulation signals ($|\Delta$$V_\text{Hanle}|$) with increasing temperature, observed in Fig. 2(b). Such qualitative consideration has already been explained in terms of the change in the spin-detection sensitivity at the tunnel contact.\cite{Lou,Jansen2,Smith,Crooker} We emphasize that the tuning of $V_\text{Int.}$ ($\mu_\text{Int.}$) is a key to detect $\Delta$$\mu$ induced by the spin injection. 

We finally comment on the difference in the actual temperature evolution of the decrease in $|\Delta$$V_\text{Hanle}|$ between spin-extraction ($I >$ 0) and spin-injection ($I <$ 0) conditions for the large currents of $I =$ $\pm$20 $\mu$A [see Figs. 2(a) and (b)]. As mentioned above, the temperature dependent $|\Delta$$V_\text{Hanle}|$ for $I >$ 0 can be explained mainly by the change in $\rho_\text{Si}$. On the other hand, for $I <$ 0, we should consider not only the contribution of $\rho_\text{Si}$ and $\lambda_{\rm Si}$ to $|\Delta$$V_\text{Hanle}|$ but also that of the spin-detection sensitivity to $|\Delta$$V_\text{Hanle}|$. Though $|\Delta$$V_\text{Hanle}|$ is reduced with decrease in $\rho_\text{Si}$ and $\lambda_{\rm Si}$, the sensitivity of the spin detection is enhanced in a low current bias regime at the same time. This opposite feature causes the relatively gradual decrease in $|\Delta$$V_\text{Hanle}|$ at $I =$ -20 $\mu$A compared with that at $I =$ +20 $\mu$A, as shown in Fig. 2. In our opinion, it is very important for room-temperature device operation\cite{Ando3} to understand the temperature-dependent sensitivity of the spin detection at the Schottky-tunnel-barrier contact in devices with a nondegenerated Si channel. 

\section{Conclusion}  
We have studied temperature evolution of spin accumulation signals obtained by the three-terminal Hanle effect measurements for the device with a nondegenerated Si channel. The clear difference in the temperature-dependent spin signals between spin-extraction and spin-injection conditions was seen. We found that it is important for the consideration of the temperature evolution to understand not only the mechanism based on the spin diffusion model but also the sensitivity of the spin detection at the Schottky-tunnel-barrier contact. These results are also important to enhance the spin signals in the device applications at room temperature. 

\begin{acknowledgments} 
This work was partly supported by PRESTO-JST and STARC. Three of the authors (Y.A. K.K. and S.Y.) acknowledge JSPS Research Fellowships for Young Scientists. 
 \end{acknowledgments}


\begin{thebibliography}{11}
\bibitem{Dery1}
S. A. Wolf, D. D. Awschalom, R. A. Buhrman, J. M. Daughton, S. v. Moln\'ar, M. L. Roukes, A. Y. Chtchelkanova, and D. M. Treger, Science {\bf 294}, 1488 (2001); H. Dery, P. Dalal, \L. Cywi\'nski, and L. J. Sham, Nature (London) {\bf 447}, 573 (2007).
\bibitem{Lou}
X. Lou, C. Adelmann, M. Furis, S. A. Crooker, C. J. Palmstr\o m, and P. A. Crowell, Phys. Rev. Lett. {\bf 96}, 176603 (2006); X. Lou, C. Adelmann, S. A. Crooker, E. S. Garlid, J. Zhang, S. M. Reddy, S. D. Flexner, C. J. Palmstr\o m, and P. A. Crowell, Nat. Phys. {\bf 3}, 197 (2007).
\bibitem{Ciorga}
M. Ciorga, A. Einwanger, U. Wurstbauer, D. Schuh, W. Wegscheider, and D. Weiss, Phys. Rev. B {\bf 79}, 165321 (2009).
\bibitem{Salis}
G. Salis, A. Fuhrer, and S. F. Alvarado, Phys. Rev. B {\bf 80}, 115332 (2009); G. Salis, S. F. Alvarado, and A. Fuhrer, Phys. Rev. B {\bf 84}, 041307(R) (2011).
\bibitem{Uemura}
Very recently, room-temperature detection of spin transport was demonstrated even for GaAs-based devices; T. Uemura, T. Akiho, M. Harada, K. Matsuda, and M. Yamamoto, Appl. Phys. Lett. {\bf 99}, 082108 (2011).
\bibitem{Kikkawa}
J. M. Kikkawa and D. D. Awschalom, Phys. Rev. Lett. {\bf 80}, 4313 (1998).
\bibitem{Min}
B. C. Min, K. Motohashi, C. Lodder, and R. Jansen, Nat. Mater. {\bf 5}, 817 (2006).
\bibitem{Igor1}
I. \v{Z}utic,  J. Fabian, S. C. Erwin, Phys. Rev. Lett. {\bf 97}, 026602 (2006).
\bibitem{Appelbaum}
I. Appelbaum, B. Huang, and D. J. Monsma, Nature {\bf 447}, 295 (2007).
\bibitem{Jonker1}
O. M. J. van't Erve, A. T. Hanbicki, M. Holub, C. H. Li, C. Awo-Affouda, P. E. Thompson, and B. T. Jonker, Appl. Phys. Lett. {\bf 91}, 212109 (2007).
\bibitem{Ando}
Y. Ando, K. Hamaya, K. Kasahara, Y. Kishi, K. Ueda, K. Sawano, T. Sadoh, and M. Miyao, Appl. Phys. Lett. {\bf 94}, 182105 (2009). 
\bibitem{Sasaki1}
T. Sasaki, T. Oikawa, T. Suzuki, M. Shiraishi, Y. Suzuki, and K. Noguchi, IEEE Trans. Mag. {\bf 46}, 1436 (2010).
\bibitem{Jansen}
S. P. Dash, S. Sharma, R. S. Patel1, M. P. Jong, and R. Jansen, Nature (London) {\bf 462}, 491 (2009). 
\bibitem{Suzuki}
T. Suzuki, T. Sasaki, T. Oikawa, M. Shiraishi, Y. Suzuki, and K. Noguchi, Appl. Phys. Express {\bf 4}, 023003 (2011). 
\bibitem{Ando3}
Y. Ando, Y. Maeda, K. Kasahara, S. Yamada, K. Masaki, Y. Hoshi, K. Sawano, K. Izunome, A. Sakai, M. Miyao, and K. Hamaya, Appl. Phys. Lett. {\bf 99}, 132511 (2011).
\bibitem{Appelbaum2}
B. Huang, D. J. Monsma, and I. Appelbaum, Phys. Rev. Lett. {\bf 99}, 177209 (2007).
\bibitem{Sasaki2}
T. Sasaki, T. Oikawa, T. Suzuki, M. Shiraishi, Y. Suzuki, and K. Noguchi, Appl. Phys. Lett. {\bf 96}, 122101 (2010).
\bibitem{Maeda}
Y. Maeda, K. Hamaya, S. Yamada, Y. Ando, K. Yamane, and M. Miyao, Appl. Phys. Lett. {\bf 97}, 192501 (2010).
\bibitem{Ando2}
Y. Ando, K. Kasahara, K. Yamane, Y. Baba, Y. Maeda, Y. Hoshi, K. Sawano, M. Miyao, and K. Hamaya, Appl. Phys. Lett. {\bf 99}, 012113 (2011).
\bibitem{ref}
The influence of the electric fields on the spin injection efficiency, discussed in Phys. Rev. B {\bf 66}, 235302 (2002), can be ignored because our device has the relatively large interface resistance and there is almost no influence of the spin absorption into the CoFe contact. 

\bibitem{Fert}
A. Fert and H. Jaffr\`es, Phys. Rev. B {\bf 64}, 184420 (2001); A. Fert, J.-M. George, H. Jaffr\`es, and R. Mattana, IEEE Trans. Electron Devices {\bf 54}, 921 (2007).
\bibitem{Pearson}
G. L. Pearson and J. Bardeen, Phys. Rev. {\bf 75}, 865 (1949); F. J. Morin and J. P. Maita, Phys. Rev. {\bf 96}, 28 (1954).
\bibitem{Jansen2}
R. Jansen and B. C. Min, Phys. Rev. Lett. {\bf 99}, 246604 (2007).
\bibitem{Smith}
A. N. Chantis and D. L. Smith, Phys. Rev. B {\bf 78}, 235317 (2008).
\bibitem{Crooker}
S. A. Crooker, E. S. Garlid, A. N. Chantis, D. L. Smith, K. S. M. Reddy, Q. O. Hu, T. Kondo, C. J. Palmstr\o m, and P. A. Crowell, Phys. Rev. B {\bf 80}, 041305(R) (2009).



\end{thebibliography}
\end{document}